\documentclass[apl,superscriptaddress,reprint,amssymb,aps,floatfix,showkeys]{revtex4-1}
\usepackage{amsmath}%
\usepackage{amsfonts}%
\usepackage{amssymb}%
\usepackage{graphicx}
\usepackage{color}
\usepackage{tabularx}
\usepackage{braket}

\begin{document}
\title{Probing Light-Matter Interaction with Topological Data Analysis}

\author{Timothy Holt}
\affiliation{ARC Centre of Excellence for Dark Matter Particle Physics, Department of Physics, University of Western Australia, 35 Stirling Highway, Crawley WA 6009, Australia}

\author{Jeremy Bourhill}
\affiliation{ARC Centre of Excellence for Dark Matter Particle Physics, Department of Physics, University of Western Australia, 35 Stirling Highway, Crawley WA 6009, Australia}

\author{Maxim Goryachev}
\email{maxim.goryachev@uwa.edu.au}
\affiliation{ARC Centre of Excellence for Dark Matter Particle Physics, Department of Physics, University of Western Australia, 35 Stirling Highway, Crawley WA 6009, Australia}

\begin{abstract}

We explore application of Topological Data Analysis to study light matter interaction through scattering response data in different dimensions. This method is robust against Fano resonance backgrounds in both strong and weak coupling regimes, maintaining accuracy even with reduced mode contrast, distorted lineshape, and the introduction of random trace noise. It scales to any number of interacting modes, reflecting the system’s effective degrees of freedom. Crucially, TDA is not merely peak counting but reveals phase-encoded features in the scattering response and may be used even for a fully saturated amplitude response. The analysis is also applied to a three mode system with time reversal symmetry breaking, revealing change in apparent number of loops and voids in combined two way scattering data. This approach is demonstrated to differentiate the three Dyson ensembles through their topological complexity and probability density functions, enabling analysis of complex modal systems. 

\end{abstract}
\date{\today}
\maketitle

\section*{Introduction}

Light-matter interaction is a cornerstone of modern physics, underpinning advancements in quantum computing, sensing, and metrology \cite{LaPierre:2022aa}. This interaction facilitates precise control over quantum systems, enabling coherence and entanglement effects to be harnessed in groundbreaking technologies. 

Quantum computing, for example, utilises electromagnetic fields across the spectrum for quantum state manipulation and information processing, especially in applications such as microwave control of superconducting circuits \cite{RevModPhys.93.025005} and optical trapping of atoms \cite{Orsi:2024aa}. Additionally, in quantum sensing, interaction with light enables unparalleled sensitivity, for instance in atomic vapours used for magnetometers \cite{Fabricant:2023aa} or with magnons exploited in hybrid quantum systems \cite{Goryachev:2014ad} for detecting weak magnetic fields. Quantum metrology is yet another field that exploits these interactions, leveraging light-matter coupling in optical atomic clocks to achieve unprecedented precision in timekeeping and navigation \cite{Colombo:2022aa}.

These technologies also enable fundamental tests of physics, such as probing quantum mechanics at macroscopic scales or searching for violations of fundamental symmetries \cite{PhysRevB.89.224407,PhysRevB.93.144420}. Beyond electromagnetism, acoustic vibrations present pathways for exploring wave-matter interactions in mechanical domains, further enriching the toolkit for quantum technologies and enabling cross-disciplinary breakthroughs in modern physics. Due to the ubiquitous nature of light-matter interactions, additional applications or methods for analysis are always highly valuable.

Topological Data Analysis (TDA) is a framework for analysing complex datasets by studying their shape and structure using algebraic topology tools \cite{Chazal:2021aa}. It focuses on extracting robust, high-level features such as connected components, loops, and voids, aiming to uncover underlying patterns and relationships in data. The core TDA technique in this work is persistent homology, which tracks the appearance and persistence of these topological features. The approach involves constructing a point cloud from a set of data and connecting points within a chosen distance of each other. The map of connected points is known as a simplicial complex, and reveals key topological features in the data. The connection distance acts as a filtration parameter, and varying its value allows for a persistence diagram or barcode to be produced, showing the longevity of particular topological features as the threshold is changed.

TDA is a rapidly evolving field, and it has found applications across diverse sectors such as identifying protein structure and function relationships in biology \cite{https://doi.org/10.1002/cnm.2914}, detecting market anomalies in financial systems \cite{10.1145/3745533}, and analysing neural activity in brain networks \cite{5872535}. Its ability to handle high-dimensional, noisy, and unstructured data makes it a powerful tool in machine learning, where it is used for feature extraction and dimensionality reduction \cite{8999144}. Recent applications have been expanded to various fields of physics, such as for generally classifying the magnetic field lines of specific configurations \cite{BOHLSEN2025134595}, characterising band structures of optical media \cite{10.1063/5.0041084}, or analysing chaos in open quantum systems \cite{PhysRevE.107.044204}.

In this work, we apply elements of TDA to analyse light-matter interactions as captured by scattering parameters (S-parameters) \cite{Kurokawa:1965aa,Choma:2009aa}. These values are derived from a system's scattering matrix, and describe the electrical behaviour of linear microwave and radio frequency networks by quantifying how signals are transmitted and reflected at each port, enabling the simulation and analysis of system dynamics over a wide frequency range. In particular, we employ the transmission coefficient $S_{21}(\omega)$ on the complex plane, which is the complex ratio of power received at port 2 over the power incident from port 1. By leveraging the $\mathrm{H}_0$, $\mathrm{H}_1$, and $\mathrm{H}_2$ topological features (connected components, loops, and voids respectively) in the data, we extract robust insights into the underlying dynamics of the system. Persistent homology provides a systematic way to quantify these features across different scales, offering a novel perspective on scattering behaviour and its relationship to the physical properties of the system. This approach provides a deeper understanding of the  structure present in $S_{21}$ measurements and highlights the utility of TDA in analysing complex experimental data.

Section \ref{sec:1} outlines the basic light-matter interaction model involved in this work, considering how the number of $\mathrm{H}_1$ loops found by persistent homology methods aides in the identification of mode interactions. Section \ref{sec:2} follows by considering Avoided Level Crossings (ALCs), using persistent homology to identify interactions from a higher dimensional picture of a dataset. Next, Section \ref{sec:3} analyses relevant methods when realistic trace noise is introduced, then Section \ref{sec:4} examines the original model when the system response becomes nonreciprocal and additional modes are introduced. Section \ref{sec:5} further explores these methods for complex systems involving high degrees of freedom, using TDA techniques to differentiate between Dyson ensembles. The concluding section explores additional applications for TDA and persistent homology in relevant fields, and reviews the full picture of results in this work.

\section{Spin Ensemble Coupled to a Cavity}\label{sec:1}

This section considers the light-matter interaction between a spin ensemble and a cavity, modelling the system as a linear combination of two coupled harmonic oscillators. This is the basic model that forms the foundation for the subsequent sections in this work. The frequency dependent scattering coefficient $S_{21}(\omega)$ is constructed by calculating the bare ideal response $\widetilde{S}_{21}(\omega)$ and introducing a Fano distortion baseline,

\begin{equation}\label{FF0001G}
    \begin{aligned}
        \widetilde{S}_{21}(\omega)&=1-\frac{\kappa}{j(\omega-\omega_c)- \frac{ \gamma}{2} + \frac{g^2}{j(\omega-\omega_a)-\frac{\Gamma}{2}}},\\S_{21}(\omega)&=\frac{\widetilde{S}_{21}(\omega)+be^{j\phi}}{1+be^{j\phi}}
    \end{aligned}
\end{equation}

where $\kappa$ is the input and output coupling constant of photons to the cavity (assumed to be symmetric at ports 1 and 2), $g$ is the coupling constant between light and matter, $\omega_a$ is a tuneable (e.g. with magnetic field) matter resonance, $\gamma$ and $\Gamma$ characterise decay rate in both light and matter parts, and $b$ and $\phi$ are coefficients of Fano distortion. Note that these equations characterise a system exhibiting an ALC subject to a Fano resonance. Typically, ALCs represent light-matter interaction points that can be quantified by a cooperativity $\mathcal{C}$. This constant is introduced as $\mathcal{C} = g^2/{\gamma \Gamma}$ and classified into strong ($\mathcal{C}>1$) and weak ($\mathcal{C}<1$) regimes. 

\begin{figure}[h!]
     \begin{center}
            \includegraphics[width=0.49\textwidth]{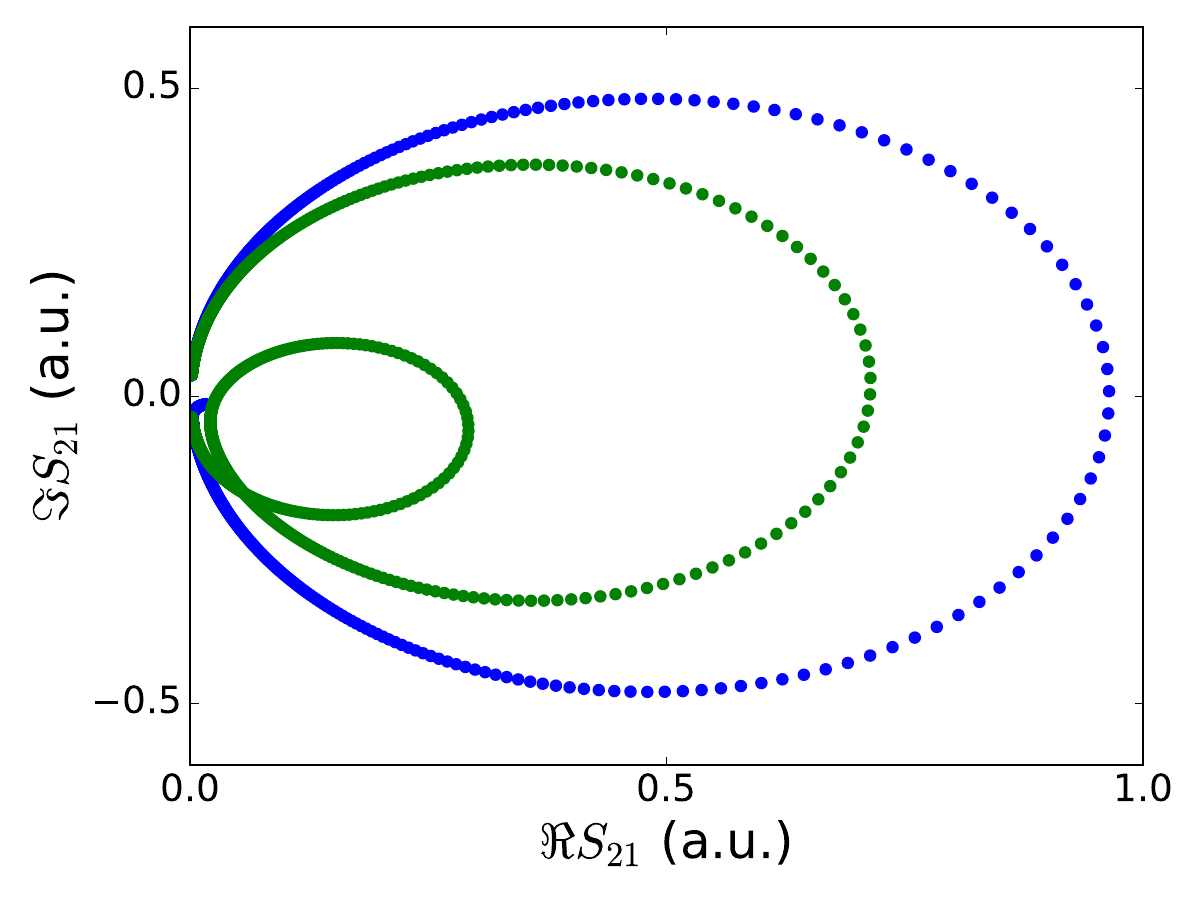}
            \end{center}
    \caption{Real and Imaginary parts of scattering coefficients $S_{21}$ form point clouds for completely detuned (blue) and fully coupled (green) systems.}%
   \label{loops-1}
\end{figure}

As a preparation step for TDA in this paper, we construct a point cloud in 2D by decomposing Equation \ref{FF0001G} into $S_{21}(\omega) = X(\omega)+jY(\omega)$ for a given system with fixed $\omega_a$, $g$, $\gamma$ and $\Gamma$. The cloud is a range of points $(X(\omega_i),Y(\omega_i))$ for $1601$ $\omega_i$ values taken in an adjustable range $[\omega_1,\omega_2]$ encompassing the resonance frequency $\omega_0$. For all analysis purposes, all frequency labels $\omega_i$ and orderings with respect to frequency are dropped and all points are treated as point coordinates on the $X-Y$ plane. Examples of the raw data point cloud for a system in the coupled ($\omega\sim\omega_c$ and $g\ne0$) and uncoupled ($g=0$ or $|\omega_a-\omega_c|\gg g$) cases are shown in Fig.~\ref{loops-1}. This result demonstrates the existence of one and two $\mathrm{H}_1$ loops in the data, highlighting the ability to determine the system's degree of coupling purely via TDA.

In this work we employ {\tt scikit-tda/Ripser} \cite{ctralie2018ripser} for all TDA calculations. The package's {\tt ripser(R)[`dgms']} function computes the persistent homology of a point cloud  $R$, represented as an array of 2D points on the $X-Y$ plane. It outputs a list of persistence diagrams, where each diagram corresponds to a homology dimension ($H_0$ for connected components,  $H_1$  for loops, etc.). Each diagram also contains intervals $[b, d]$ that represent the birth $b$ and death $d$ scales of topological features in that dimension. We can introduce persistence or lifespan of each loop as $\Delta \tau = d-b$. These intervals capture the evolution of the dataset’s topological structure across different scales, providing insights into its geometric and topological properties. This method is widely used in TDA to identify significant patterns and structures in data.

\begin{figure*}[ht]
     \begin{center}
            \includegraphics[width=0.99\textwidth, angle=180]{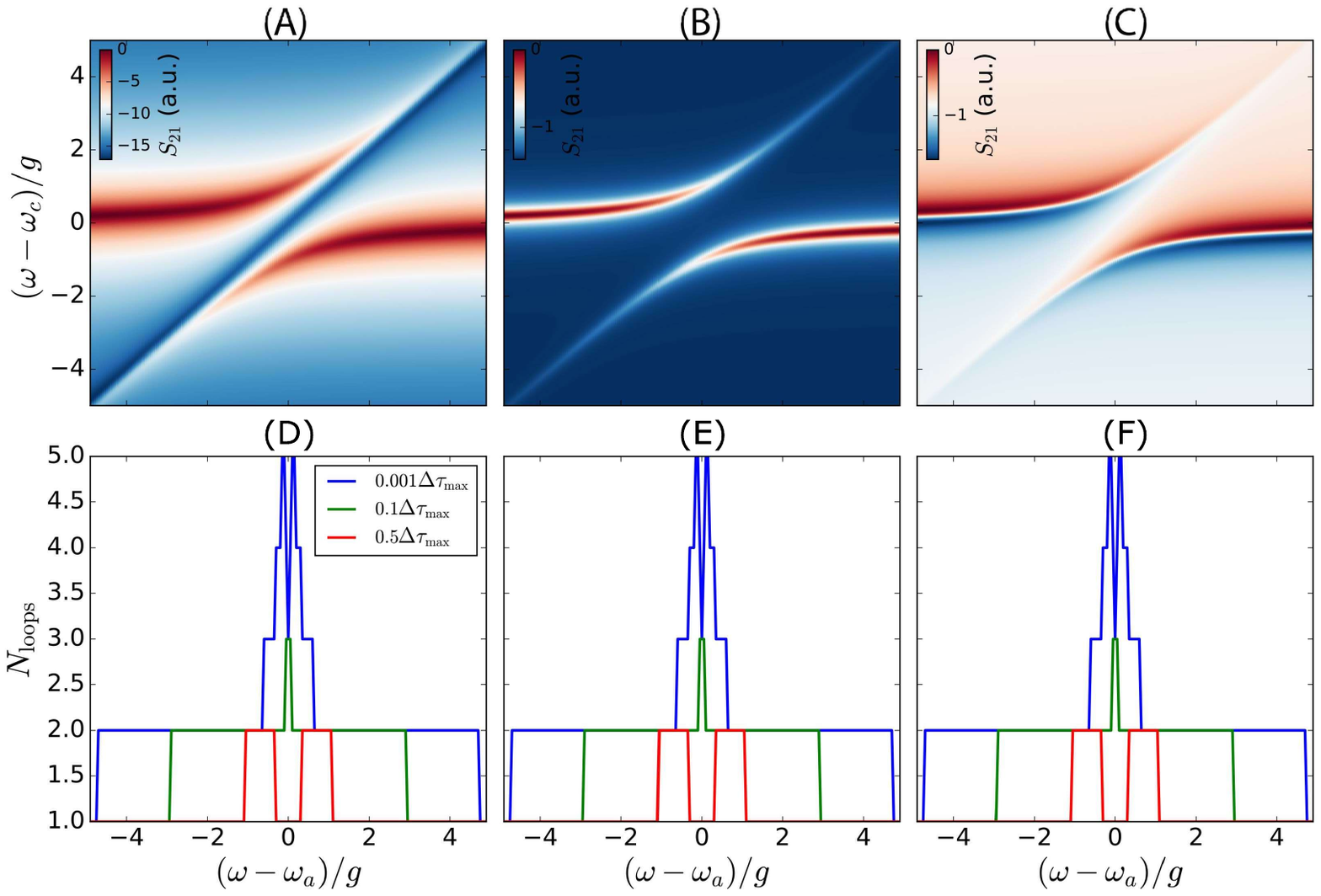}
            \end{center}
    \caption{ (A)-(C): Scattering coefficients $S_{21}$ of a light-matter interaction system demonstrating ALCs in the strong coupling regime as a function of cavity and matter detunings: (A) no Fano resonance $b=0$, (B) Fano distortion with $b = 5\times10^4$, $\phi=0$, (C) Fano distortion with $b = 5\times10^4$, $\phi=\pi/2$. (D)-(F): respective number of loops for 3 different values of threshold: $0.001 \tau_\mathrm{max}$ (blue),  $0.1 \tau_\mathrm{max}$ (green) and  $0.5 \tau_\mathrm{max}$ (red).}%
   \label{strong-1}
\end{figure*}

Fig.~\ref{strong-1} shows the basic interaction system in the strong coupling regime, where subplots (A)-(C) demonstrate typical ALCs for the three choices of Fano parameters. In this regime $\Gamma = \gamma$ and $g = 3\Gamma$ (or $\mathcal{C} = 9$), and the three Fano configurations are no Fano resonance $b=0$, Fano distortion with $b = 5\times10^4$ and $\phi=0$, and Fano distortion with $b = 5\times10^4$ and $\phi=\pi/2$. The parameter $b$ for the second case is taken large enough to place the system in a regime where the resonance peak is significantly obscured. In the third configuration, $\phi$ is adjusted to produce a distorted resonance profile characterised by both positive and negative extrema. Comparing these three cases, it is observed that the Fano resonance reduces the contrast of the modes, and in the case of nonzero phase also introduces asymmetry of each peak. 

Subplots (D)-(F) demonstrate results of the TDA analysis in the form of number of loops in the 2D data for the three cases. In each case, three curves represent the loop threshold size $\Delta \tau_\mathrm{th}$ in terms of maximum persistence value $\Delta \tau_\mathrm{max} = \max(b-d)$ for each set of data. 

\begin{figure*}[ht!]
     \begin{center}
            \includegraphics[width=0.99\textwidth, angle=180]{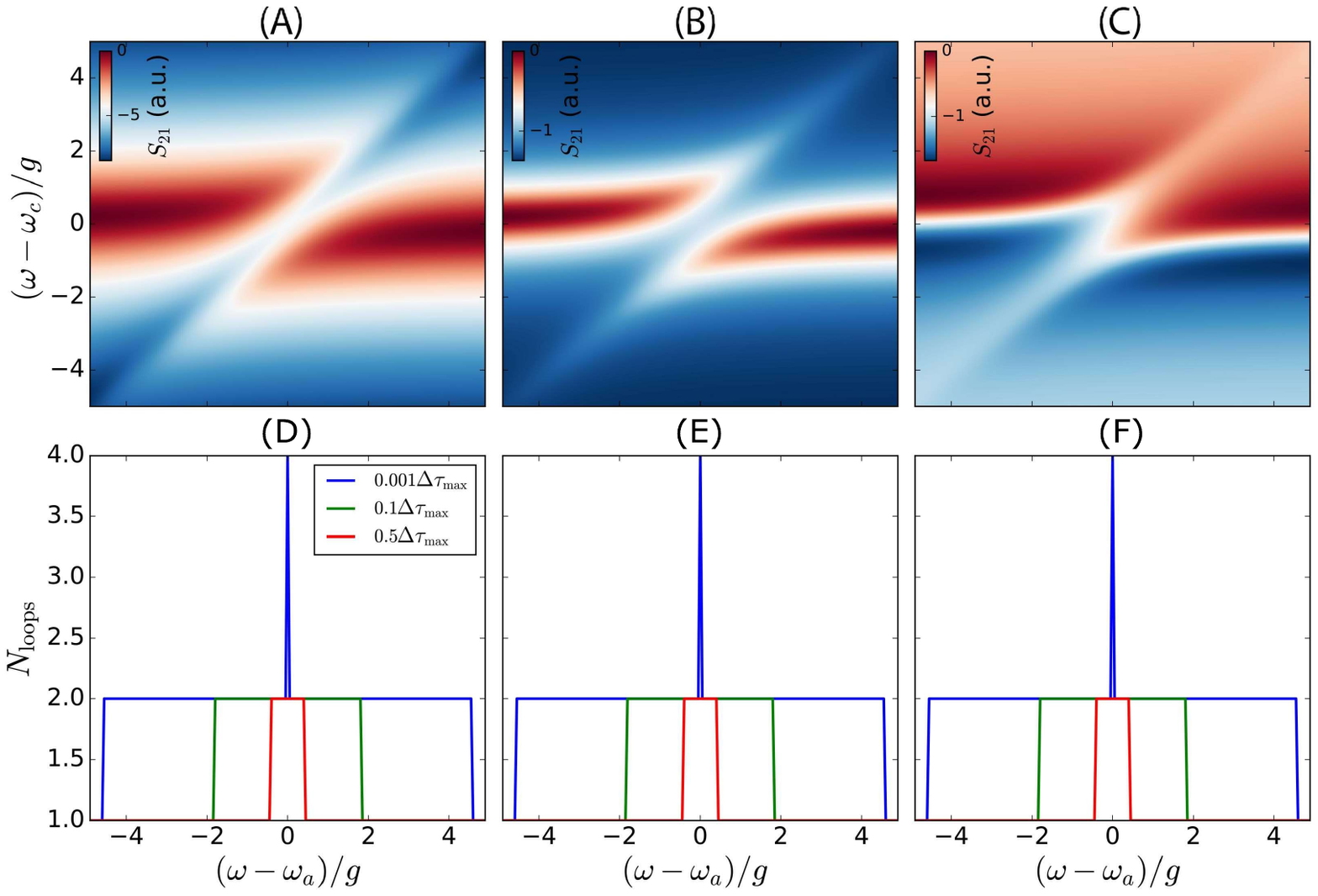}
            \end{center}
    \caption{ (A)-(C): Scattering coefficients $S_{21}$ of a light-matter interaction system demonstrating ALCs in the weak coupling regime as a function of cavity and matter detunings: (A) no Fano resonance $b=0$, (B) Fano distortion with $b = 5\times10^4$, $\phi=0$, (C) Fano distortion with $b = 5\times10^4$, $\phi=\pi/2$. (D)-(F): respective number of loops for 3 different values of threshold: $0.001 \tau_\mathrm{max}$ (blue),  $0.1 \tau_\mathrm{max}$ (green) and  $0.5 \tau_\mathrm{max}$ (red).}
   \label{weak-1}
\end{figure*}

For all thresholds the minimum number of loops is one, corresponding to a single photonic resonance that is coupled to and observed directly irrespective of matter resonance detuning. Near the interaction point, the number of loops grows reflecting effective appearance of higher order components in the system, i.e. coupling to the matter degree of freedom. The number of loops with threshold of $0.5\Delta \tau_\mathrm{max}$ grows to 2 near the ALC where two resonant peaks in a single $S_{21}$ spectra could be observed. In this way, the TDA detects the number of degrees of freedom in the system near the corresponding ALC. These two loops could be seen in point clouds shown in Fig.~\ref{loops-1}. By comparing different Fano parameters, it is clear from this plot that $N_\mathrm{loop}$ is independent of the parameters of the Fano background. 

Fig.~\ref{weak-1} shows the simulated system in the weak coupling regime, where the parameter choice $g=2\Gamma/3$ or $\mathcal{C} = 4/9$ for the same three cases with respect to the Fano resonance is analysed. The result suggests that the TDA in this case is still able to detect interaction through the number of loops. Similarly to the response of the strongly coupled system, subplots (D)-(F) indicate that the TDA result is independent of the Fano parameters. 

\begin{figure}[h!]
     \begin{center}
            \includegraphics[width=0.49\textwidth]{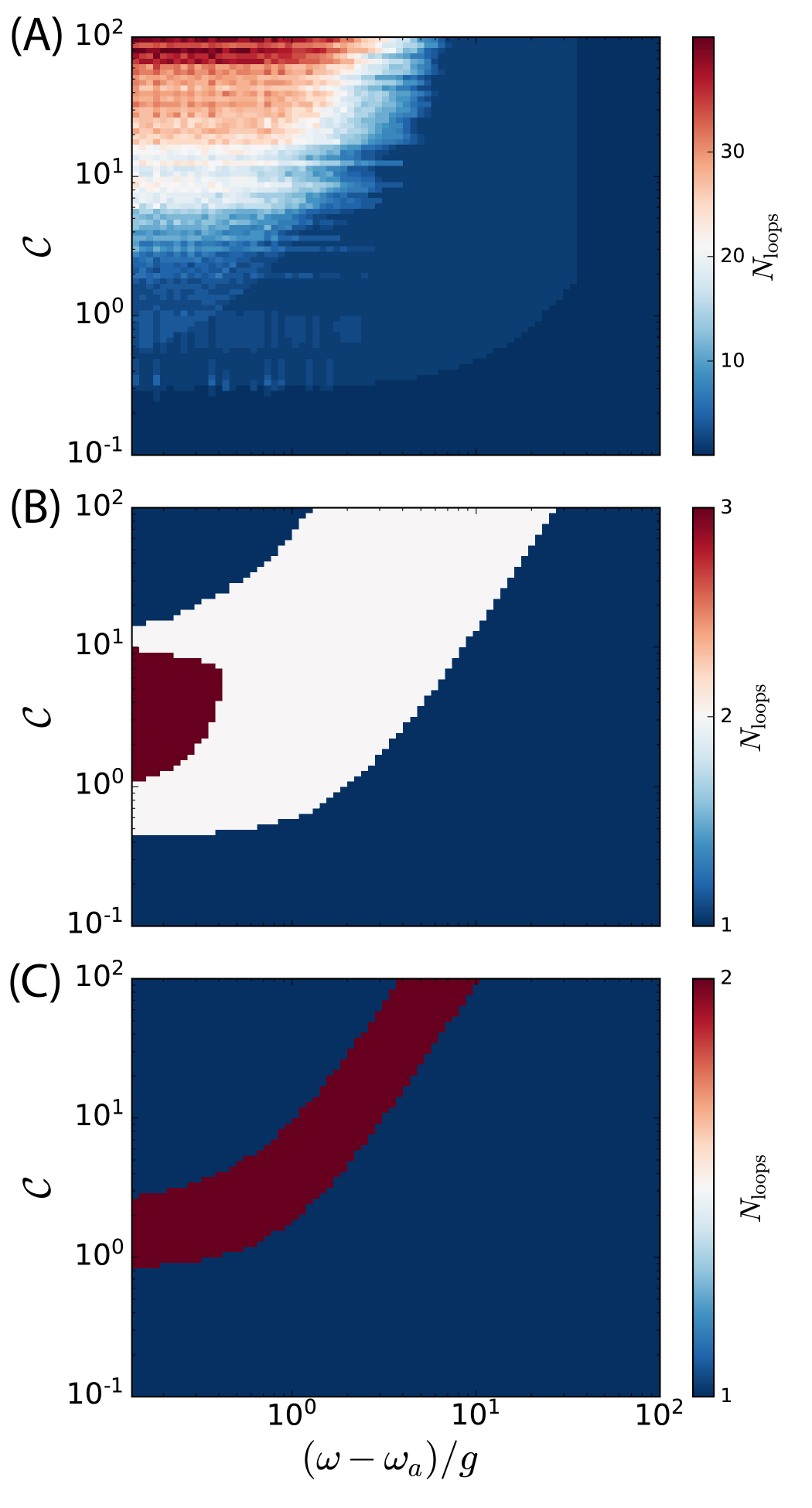}
            \end{center}
    \caption{Number of loops for 3 different values of threshold as a function of cooperativity $\mathcal{C}$: (A) $0.001 \tau_\mathrm{max}$, (B) $0.1 \tau_\mathrm{max}$ and  (C) $0.5 \tau_\mathrm{max}$.}%
   \label{coupling-1}
\end{figure}

To observe the minimum level of light-matter coupling this approach can detect, we calculate system response as a function of cooperativity $\mathcal{C}$ and matter detuning $(\omega-\omega_a)/g$ in logarithmic units. The result is shown in Fig.~\ref{coupling-1}, showing that the interaction becomes detectable for cooperativities in the range 0.1 to 1. For all parameter values, the $0.5 \tau_\mathrm{max}$ case shows discrete values 1 and 2. At the same time, lower threshold values allow the detection of the interaction at significantly lower values of $\mathcal{C}$ as seen from comparison of (A) and (B).

\begin{figure}[th!]
     \begin{center}
            \includegraphics[width=0.49\textwidth]{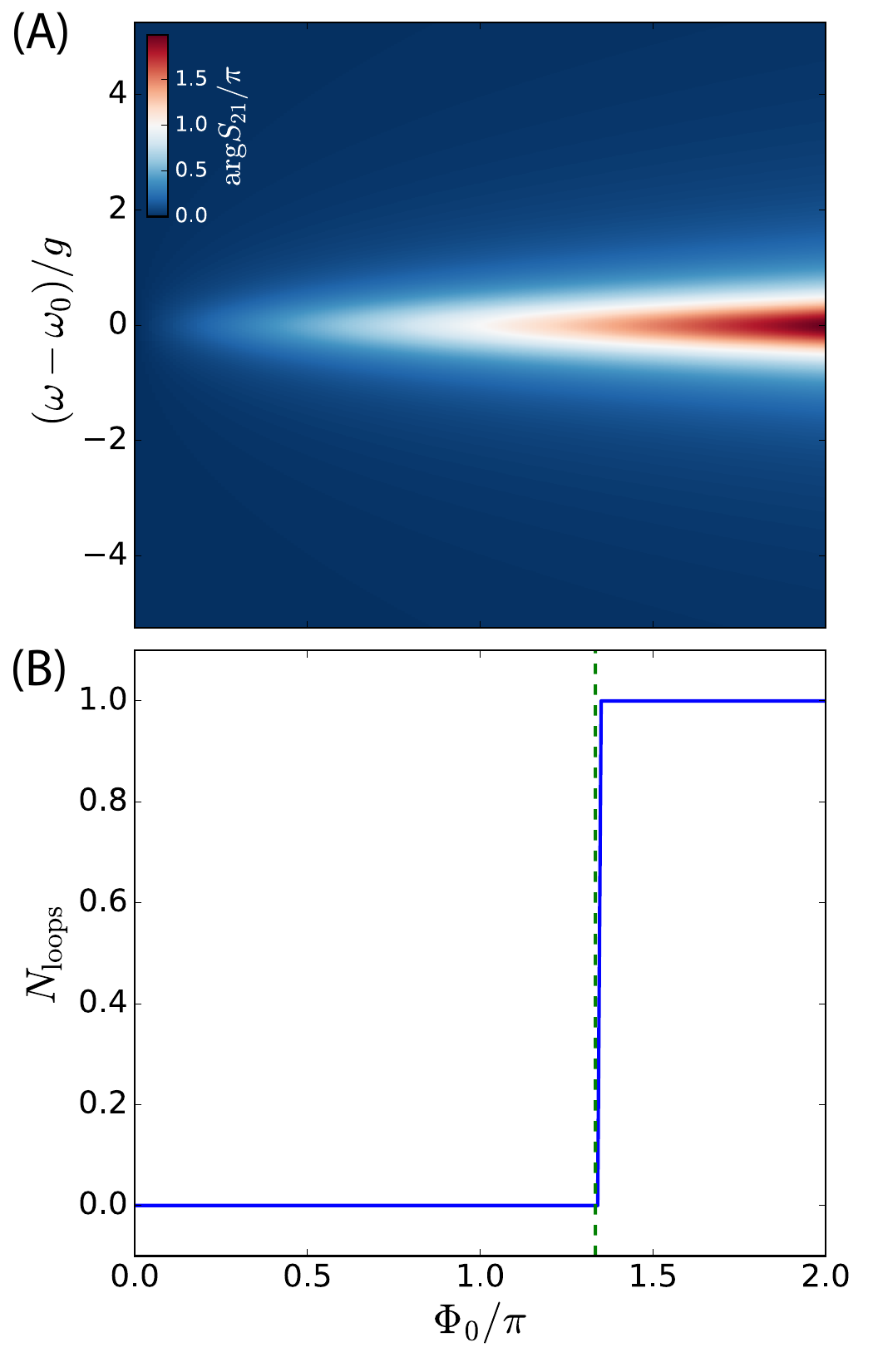}
            \end{center}
    \caption{(A) $\arg S_{21}$ response of the all pass filter defined in Eq.\ref{FF0002G} for a range of $\Phi_0$. (B) Corresponding number of loops identified using TDA. Vertical dashed line shows $\Phi_0=4/3\pi$.}
   \label{allpass-1}
\end{figure}

For the simple cases presented in this section, each TDA result can be alternatively achieved by looking at the resonant peaks in the $S_{21}$ magnitude data to observe the interaction. This however becomes far more difficult for higher order or very noisy interacting systems, and so TDA methods can become more useful. To demonstrate the inequivalence between amplitude detection and TDA of scattering response, we explore an all pass filter with a Lorentzian resonance in the phase space,

\begin{equation}
	\label{FF0002G}
	{S}_{21}(\omega) = \exp\Big[ j\frac{\Phi_0}{1 + ((\omega - \omega_0) / (\Delta / 2))^2}\Big],
\end{equation} 

where $\omega_0$ is the central resonance frequency, $\Delta$ is corresponding bandwidth, and $\Phi_0$ is corresponding maximum phase deviation. Since $|S_{21}(\omega)|=1$ for all $\omega$, this feature cannot be detected by simply looking at the system amplitude response, making an alternate analysis method necessary. The corresponding system response for different values of $\Phi_0$ is shown in Fig.~\ref{allpass-1}. This simulation reveals that for $\Phi_0>4\pi/3$, the feature is found as a single topological loop in the scattering response. This feature would be undetectable if only amplitude data was considered.

\section{Cooperativity Dependence}\label{sec:2}

When considering one-dimensional $\mathrm{H}_1$ loops, the above methods counted using individual $S_{ij}(\omega)$ scattering traces as a function of resonance frequency $\omega_a$. In addition to this, TDA can be used to probe ALCs as single point clouds. The number of loops per ALC can be quantified for the chosen two-parameter scattering manifold, producing a top-down picture of the data in a higher dimensional space. To demonstrate this idea, we compute $N_\mathrm{loops}$ for the transmission response $S_{21}(\omega,\omega_a)$ while varying cooperativity $\mathcal{C}$, thereby revealing how the topology of the scattering manifold evolves across coupling regimes. We employ a randomised sampling grid in $(\omega,\omega_a)$ to suppress aliasing and minimise sampling artefacts.

\begin{figure}[h!]
     \begin{center}
            \includegraphics[width=0.49\textwidth]{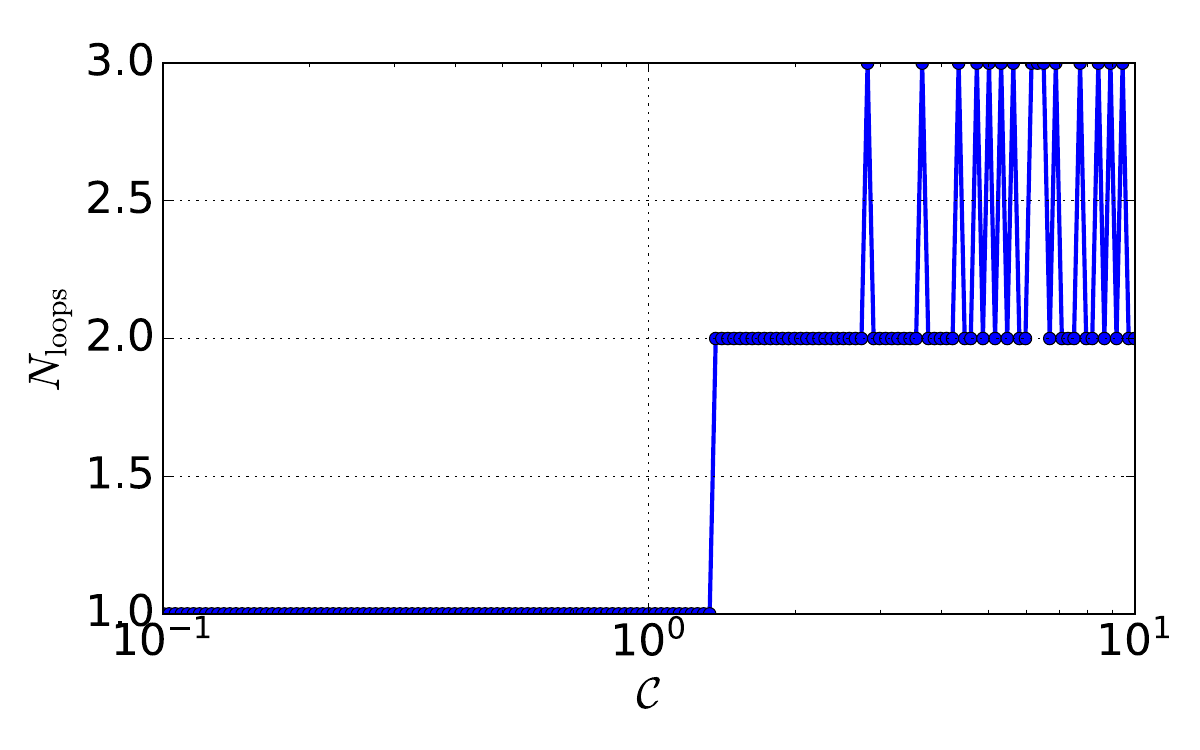}
            \end{center}
    \caption{Number of loops in $(\Re S_{ij}(\omega,\omega_a),\,\Im S_{ij}(\omega,\omega_a))$) cloud of points of ALCs as a function of cooperativity $\mathcal{C}$.}%
   \label{ALCloops-1}
\end{figure}

The results of this process are shown in Fig.~\ref{ALCloops-1}. This figure illustrates the transition from weak to strong coupling near $\mathcal{C}\approx 1$. The transition point is expected to be at unity but the simulation is limited by the chosen number of points, which must be kept low due to the sizeable parameter space. For $\mathcal{C}<1$, the point cloud yields a single persistent $H_1$ loop; all traces are effectively identical and form one cycle. As $\mathcal{C}$ exceeds unity, two distinct loop classes emerge, consistent with mode hybridisation and the onset of an ALC as explored previously. 

At larger $\mathcal{C}$ values, the estimated loop count shows increased variability, which arises due to under sampling issues. The simulation result is required to both resolve the resonance line shapes and span a wide frequency window using a finite number of points. This contradiction results in a fixed sampling budget and hence a reduced effective resolution. Ultimately the higher-dimensional approach is seen to be an efficient method for considering a system exhibiting ALC behaviour.

\section{Trace Noise Analysis}\label{sec:3}

Another important consideration for this work is how introducing random trace noise to the otherwise ideal Fano-distorted dataset affects the accuracy of TDA techniques. Noise-resistant techniques are incredibly valuable due to their applicability to real systems and data. By randomly skewing each $S_{21}$ point with a normal distribution, a good approximation of an unknown background interference is introduced on top of the known Fano background. The magnitude of this noise is specified by a Signal to Noise Ratio (SNR), which can be used with the peak-to-peak signal to calculate a noise standard deviation value $\sigma$. Visually, the noise produces graphs similar to Fig.~\ref{loops-1} but with increasing deviation from the loop shape as the SNR is reduced. 

\begin{figure}[t!]
     \begin{center}
            \includegraphics[width=0.49\textwidth]{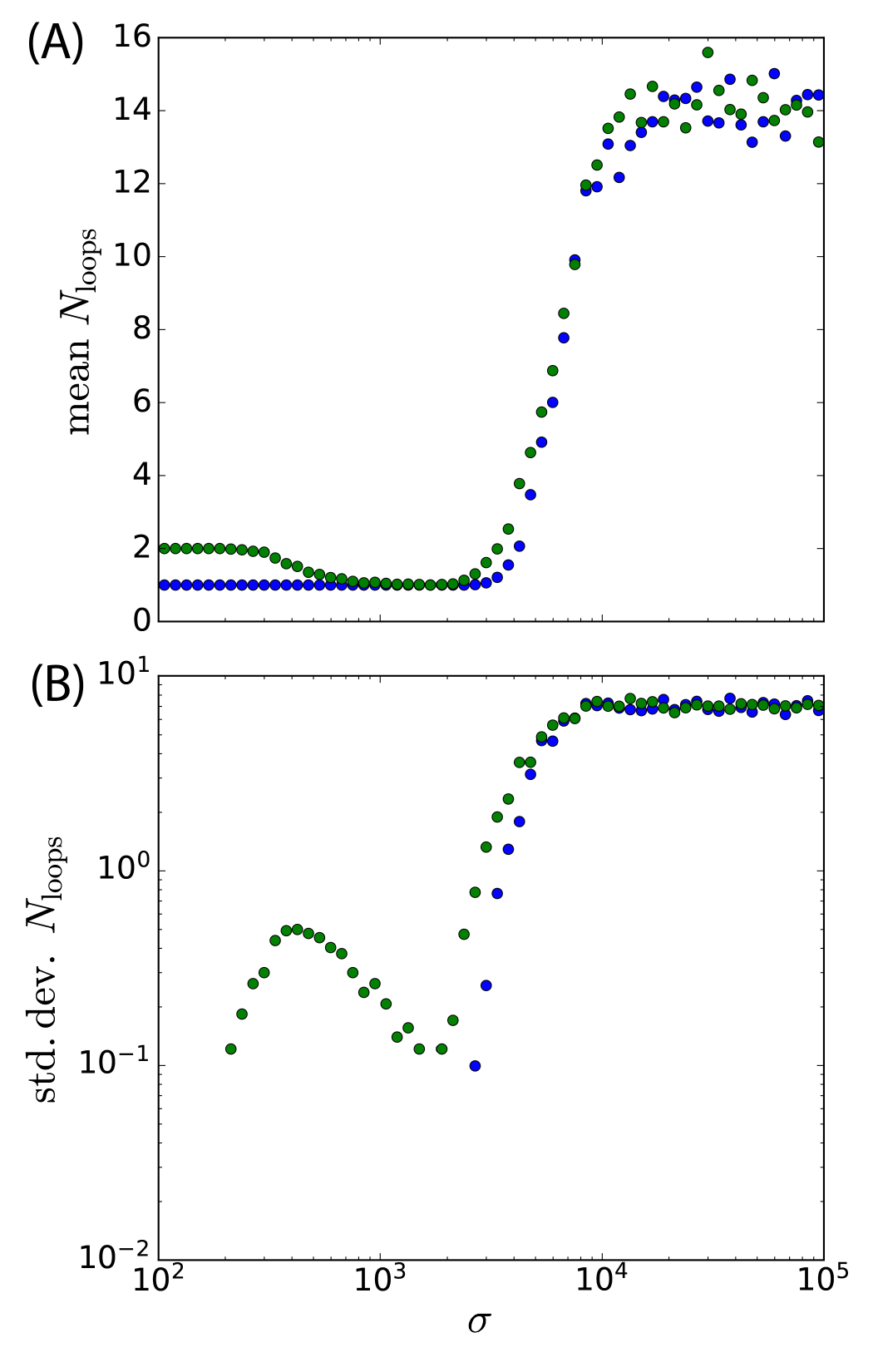}
            \end{center}
    \caption{(A) The mean and (B) standard deviation of the counted number of loops compared to the standard deviation of the applied random trace noise $\sigma$ using a detection threshold of $0.5 \tau_\mathrm{max}$. Both the completely detuned (blue) and fully coupled (green) systems are plotted.}%
   \label{loops-2}
\end{figure}

It is useful to analyse the levels of noise required before TDA becomes impossible in order to observe the limitations of this work. Using the established {\tt ripser} methods, loop counting is performed on an SNR range corresponding to a $\sigma$ range of $10^2$ to $10^5$. Fig.~\ref{loops-2} displays the results of this analysis, comparing the mean and the standard deviation of the loop count $N_{loops}$ at each specified trace noise $\sigma$. The blue and green plots correspond to the detuned and coupled system states in Fig.~\ref{loops-1} respectively, where the noiseless blue data resolves to one loop and the noiseless green data resolves to two. These results were found using the threshold $0.5 \tau_\mathrm{max}$ to produce the most sensitivity to the loops.A value for $\sigma$ is produced using a standard SNR relation,

\begin{equation}
	\label{SNRsigma}
	\sigma = \frac{\Delta S_{21}}{10^{SNR/20}},
\end{equation} 

where $\Delta S_{21}$ is the peak to peak change in the $S_{21}$ signal, and $SNR$ is the desired Signal to Noise Ratio for the simulated system state.

In Fig.~\ref{loops-2}, TDA resolves a single loop in the data below $\sigma=3\times10^3$. Above this, $N_{loops}$ diverges and eventually plateaus to a value indicative of the threshold being used. For the double loop trace, the second loop is only resolved completely near $\sigma=2\times10^2$. Additionally, the standard deviation of $N_{loops}$ reduces from a baseline to zero near $\sigma=2\times10^3$. When considering a second loop however, the standard deviation quickly rises again then vanishes when the second loop is fully resolved. The trend in standard deviation appears to act as an earlier indication of an interaction than the mean, enabling this method to be further resistant to a weak SNR.

\section{System with Broken $\mathcal{T}$-Symmetry}\label{sec:4}

The results of Sections \ref{sec:1}, \ref{sec:2}, and \ref{sec:3} explore the application of TDA to two dimensional data from single $S_{21}$ measurement sets. This implies a two port device with unbroken time ($\mathcal{T}$) symmetry, and that the system response is reciprocal (i.e. $S_{21}=S_{12}$). Given that $S_{11}$ produces the same information up to a simple transformation, all features of such systems can be fully confined in 2 dimensions. If the $\mathcal{T}$ symmetry is broken however, the full system response can no longer be represented solely by $S_{21}(f)= X+jY$ and $S_{12}(f)= W+jV$ must be considered. In this case, the total data lives in 4 dimensions as both scattering coefficients are complex. This implies that in addition to considering loops in 2D, one can look for $\mathrm{H}_2$ voids (the higher dimensional loop equivalent) in 4D. Overall, one can analyse 3 data sets: $(X,Y)$, $(W,V)$ and $(X,Y,W,V)$, where for the latter set it is possible to search for both loops and voids.

Additionally, such analysis can scale above the simple case of a two mode system. To illustrate this, we consider the case of a three mode interaction with clear time reversal symmetry breaking \cite{Gardin:2023aa,Bourhill:2023aa,Zhang:2020aa}. Physically this system can represent an experimentally relevant situation with a tuneable matter/spin transition and a Whispering Gallery Mode (WGM) doublet. WGMs are typically supported by photonic systems with rotational symmetry (such as cylinders, toroids, or disks), and a small deviation from this symmetry breaks degeneracy between sinusoidal and cosinusoidal solutions to the system creating a two-peaked structure characteristic of a doublet. This feature can be modelled as a two-degenerate harmonic oscillator system (each having resonance frequency of $\omega_c$) with fixed direct coupling between them defined by the coupling constant $\lambda$. The two modes of the doublet typically have different coupling to feeding lines, i.e. $\kappa_1\neq\kappa_2$. 

When considering time reversal symmetry breaking, the spin transition couples to only one mode from the doublet pair with the coupling strength $g$. The overall scattering of such a system can be calculated as \cite{Zhang:2020aa}:
\begin{equation}
	\label{FF00d1G}
	\mathbf{S}(\omega) = \mathbf{C} + \mathbf{D}\big[ -j\omega  \mathbf{I}-\mathbf{A}\big]^{-1} \mathbf{B},
\end{equation}
where due to conservation restrictions, the matrices can be written as follows:
\begin{equation}
\mathbf{A}=\left(\begin{array}{ccc}
-j \omega_1-\frac{\kappa_1}{2} & j\lambda & j g_1 \\
j\lambda & -j \omega_2-\frac{\kappa_2}{2} & j g_2 \\
j g_1 & j g_2 & -j \omega_a-\frac{\kappa_a}{2}
\end{array}\right),
\end{equation}
\begin{equation}
\mathbf{B}=\left(\begin{array}{cc}
\sqrt{\eta \kappa_{1 e}} & \sqrt{(1-\eta) \kappa_{1 e}} e^{j \mu} \\
\sqrt{(1-\eta) \kappa_{2 e}} e^{j \nu} & \sqrt{\eta \kappa_{2 e}} e^{j(\mu+\nu+\pi)} \\
0 & 0
\end{array}\right),
\end{equation}
\begin{equation}
\mathbf{C}=\left(\begin{array}{cc}
\sqrt{1-\xi} & j \sqrt{\xi} \\
j \sqrt{\xi} & \sqrt{1-\xi}
\end{array}\right),
\end{equation}
and $\mathbf{D} = -\mathbf{C}\mathbf{B}^\dagger$. Here $g_1$ and $g_2$ are light-matter coupling rates for both photonic modes, $\kappa_x$ ($x\in \{1,2,a\}$) are decay rates for 3 modes, $\xi$ is the port cross-talk constant, $\eta$ is the ratio of the coupling rate of each individual port to the total external coupling rate, and $\mu$ and $\nu$ are phase of couplings. To simplify the model, we consider the case of no cross-talk $\xi=0$, all equal decay rates, equal coupling rates $\eta=0.5$, no direct coupling $\lambda=0$, and $\sqrt{\kappa_{2e}/\kappa_{1e}}=0.99$.

\begin{figure}
     \begin{center}
            \includegraphics[width=0.49\textwidth]{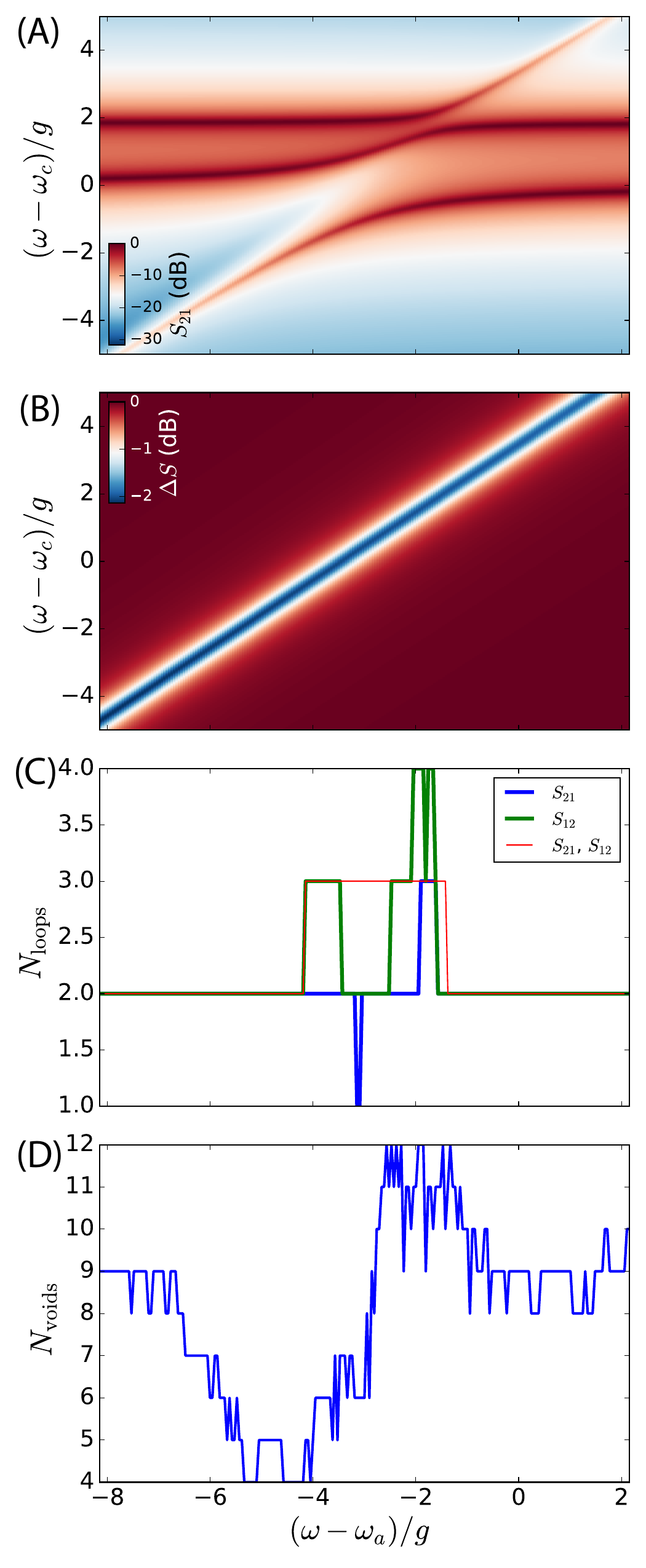}
            \end{center}
    \caption{(A) $S_{21}$ response of a doublet WGM system coupled to matter with $\mathcal{T}$ symmetry breaking as a function of matter tuning. (B) System non-reciprocity $\Delta S$. (C) Calculated number of loops in 3 data sets. (D) Calculated number of voids. }%
   \label{tripple-1}
\end{figure}

The scattering response of an example of this triple mode system for the case $\omega_2=1.0011\omega_2$, $g_1=6\Gamma$, $g_2=2\Gamma$, $\mu=-\pi/2$, and $\nu=\pi/2$ is shown in Fig.~\ref{tripple-1}. Here, subplot (A) reveals existence of 3 modes (two light and one tuneable matter) exhibiting 2 ALCs, (B) shows system non-reciprocity as $\Delta S = S_{21}-S_{12}$, (C) demonstrates the change in loop number for 3 scattering data sets near each ALC, and (D) depicts change in number of voids in the 4D data set. The result reveals the existence of 1, 2 and 3 loops in the scattering data for different values of matter detuning. The number of discovered voids changes between 4 and 12 depending on matter detuning. This plot shows some noise that depends on the number of points and frequency span. The number of points for this example is reduced to 1511 due to limitation of involved computing power, contributing to the observed noise.

This approach enables an extension to three and more port devices, where in addition to $S_{21}(\omega)$ and $S_{12}(\omega)$ one measures scattering into $n$ other ports (i.e. $S_{n1}(\omega)$ and $S_{1n}(\omega)$ for $n>2$). In this case, the data will be represented in even more dimensions. An example of this situation is for a circulator, a common microwave device with at least three ports. This device must be characterised by at least $S_{21}(\omega)$ and $S_{31}(\omega)$ measurements and their reciprocal versions to fully account for its intended purpose. As a result, the scattering data will be represented in a minimum of 4D, requiring higher order persistence calculations for similar void analysis to the 3 mode case.

\section{Complex Systems}\label{sec:5}

Complex wave-chaotic systems with a large number of overlapping modes and degrees of freedom resist detailed deterministic description, yet their statistical behaviour is governed by a few symmetry principles. In scattering experiments, the power transmission coefficient is defined using the complex transmission amplitude between two ports, i.e. $S_{21}=P_2^{out}/P_1^{in}=|T|^2$ where $T=U_2^{out}/U_1^{in}$. Over the total system, Random Matrix Theory (RMT) can be used to predict ensemble-averaged statistics for specific cases \cite{Dyson:1962aa,Wigner:1955aa}. Using RMT, the internal scattering matrix of a lossless cavity can be modelled three situations by Gaussian ensembles: the Gaussian Orthogonal Ensemble (GOE), the Gaussian Unitary Ensemble (GUE), and the Gaussian Symplectic Ensemble (GSE). Each ensemble is distinguished by its Dyson index $\beta$, counting the number of real components per element in each ensemble matrix. In this section, we simulate large ensembles of scattering trajectories from each symmetry class and show that persistent $H_1$ loop statistics obtained via TDA provides a robust, symmetry-sensitive signature that complements conventional RMT observables.

GOE ($\beta=1$) is used for the case where Time Reversal Symmetry (TRS) holds and there are no spin-orbit coupling effects. Its real/symmetric matrices predict a highly skewed transmission distribution $P(T)\propto T^{-1/2}$ and favour a small $T$, resulting in a “weak localisation” dip. When TRS is broken (e.g. by magnetised ferrites), GUE ($\beta=2$) arises and yields complex/Hermitian matrices with a uniform distribution $P(T)=1$. The variance of this distribution is half that of the GOE case. Lastly, GSE ($\beta=4$) is the most specific case and requires an effective spin-1/2 degeneracy (Kramers' symmetry). Its quaternion/Hermitian matrices enforce an alternate distribution $P(T)\propto T$, suppressing low–transmission events and yielding the smallest relative fluctuations.

These ensembles have been experimentally realised in microwave systems, with GOE-to-GUE transitions achieved by inserting nonreciprocal elements that break bosonic TRS \cite{Hul:2004aa,So:1995aa}. Achieving GSE requires engineering a pseudo-spin doublet. Two coupled degenerate modes biased with opposite nonreciprocity can enforce a $T^2 = -1$ Kramers'-type symmetry, thus realising the characteristic transmission distribution $P(T) \propto T $ \cite{Rehemanjiang:2016aa}. Characterising these ensembles provides valuable insights into the dynamics underpinning fluctuations in conductance, photon emission, and quantum state stability for a range of mesoscopic and circuit-QED systems \cite{Hemmady:2005aa}. Beyond spectral properties, RMT governs the statistical behaviour of S-matrix elements and transmission coefficients \cite{Kumar:2013aa,Savin:2003aa}. 

In complex wave-chaotic scattering experiments, the rich structure of the measured transmission and reciprocity coefficients can be revealed by treating the real and imaginary trajectories of $S_{21}(f)$ and $S_{21}(f)$ as a high-dimensional point cloud and applying persistent homology to extract its $H_1$ loops. By tallying the number of loops above a certain persistence threshold, the number of coupled modes can be determined directly from the raw scattering data similarly to previous sections. The TDA approach offers several additional distinct advantages beyond the ability to count modes without fitting to the data.

In GSE systems, each physical mode appears as a pair with Kramers' degeneracy, generating two nested loops per resonance. As a result, TDA allows the distinction of GSE systems from GOE or GUE by revealing a 2:1 loop-to-mode ratio. Additionally, by embedding both $S_{21}$ and $S_{12}$ channels one observes nearly identical loop patterns in reciprocal systems. Comparing the number, lifetimes, or total persistence of $H_{1}$ features between forward and reverse transmission directly quantifies the degree of TRS breaking, further enabling the identification of nonreciprocal GUE systems. Another advantage lies in the ability to extract ensemble-specific persistence distribution from the lifetimes of loops. The resulting distributions mirror the established RMT transmission distributions for each Gaussian ensemble, however they are extracted purely topologically without spectral averaging.

When implementing TDA in this analysis, the $N\times N$ scattering matrix is built from an effective non-Hermitian Hamiltonian $H$ for $N$ coupled resonant modes and a port-coupling matrix $\mathbf{W}$. First, one draws a $N\times N$ random Hermitian “coupling” matrix from the chosen RMT ensemble and shifts each diagonal element by $-j\kappa/2$ to represent the external linewidth $\kappa$.  This yields
$\mathbf{H} = \mathbf{H}_{\rm RMT} \;-\; \frac{j\kappa}{2}\,\mathbf{I}$.
Next, the $N\times N$ matrix $\mathbf{W}$ is constructed so that each column describes the complex amplitudes by which one of the $N$ external ports drives and extracts energy from the internal modes. The columns are normalised to $\sqrt{\kappa}$ and random phases are chosen. 

\begin{figure}[h!]
     \begin{center}
            \includegraphics[width=0.49\textwidth]{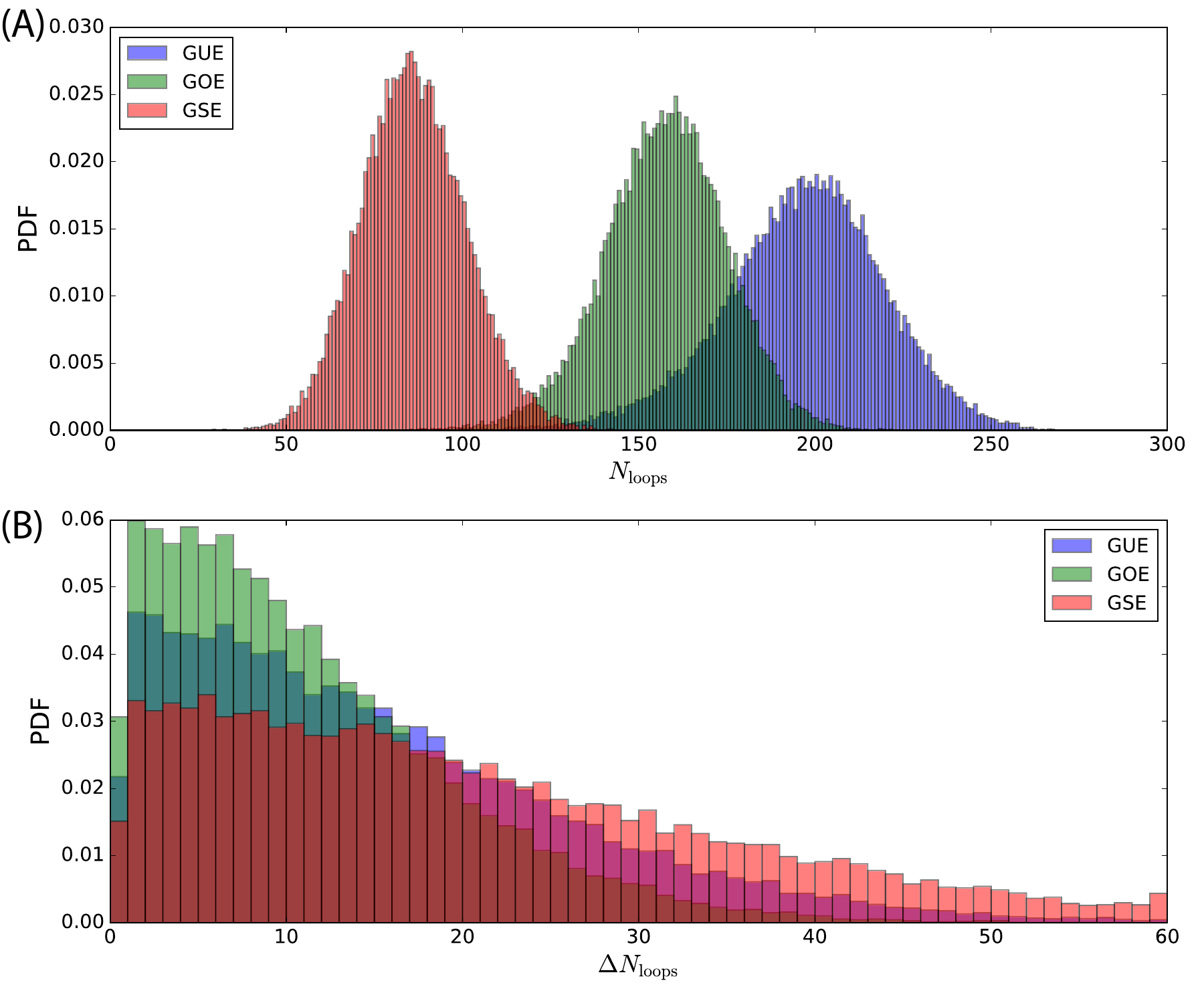}
            \end{center}
    \caption{(A) The probability density function of distribution of number of loops in 4D data for GUE, GOE and GSE system ensembles. (B). The probability density function of distribution of differences in $N_\mathrm{loops}$ forward and backward directions for these ensembles.  }%
   \label{GUEGOEGSE-1}
\end{figure}

The full frequency-dependent scattering matrix is then given by the standard input-output formula
$\mathbf{S}(\omega)\;=\;\mathbf{I} \;-\; j\,\mathbf{W}^{T}\bigl(\omega \mathbf{I} - \mathbf{H}\bigr)^{-1}\mathbf{W}\,$,
evaluated at $\omega=2\pi f$.  Finally, the entries $S_{21}(f)$ and $S_{12}(f)$ are extracted across the frequency grid to yield the forward and reverse transmission spectra. Based on these data we calculate 30,000 realisations of each class giving a distribution in the number of loops as shown in Fig.~\ref{GUEGOEGSE-1} (A). Additionally, we calculate similar distribution for the differences between loops in $S_{21}$ and $S_{12}$ shown in Fig.~\ref{GUEGOEGSE-1} (B). In both cases, loop counting is done with the threshold of $0.01 \tau_\mathrm{max}$ and 1000 total points.

In RMT the Dyson index $\beta$ not only dictates level‐spacing statistics but also controls the “roughness” of scattering trajectories and hence the topology of their loop counts.  Quantitatively, we observe that as $\beta$ increases from 1 (GOE) to 2 (GUE) to 4 (GSE), the mean number of persistent $H_{1}$ loops first rises, peaking for GUE at $\langle L\rangle\approx198$ due to maximal phase‐randomisation, and then falls sharply for GSE at $\langle L\rangle\approx85$ where strong Kramers' degeneracy enforces smooth, repulsion‐dominated curves. The variance of the loop count scales roughly inversely with $\beta$, reflecting the common convention for the Gaussian ensembles $\mathrm{Var}\sim1/\beta$ \cite{forrester2010log}. Here GUE shows the largest scatter ($\approx534$), GOE is intermediate ($\approx297$), and GSE the most self‐averaging ($\approx219$). 
%Higher‐order moments provide additional insight into ensemble‐specific tails: negative skew in GOE and GUE indicates a long tail of simple realisations (fewer loops than average), whereas GSE’s slight positive skew signals rare excursions above its low baseline. Likewise, kurtosis decreases with $\beta$, from GUE’s heavier tails ($\approx0.78$) to near‐Gaussian GSE behaviour ($\approx0.10$).  
Together, these trends illustrate how $\beta$ governs not just spectral repulsion but the full statistical and topological complexity of multiple‐scattering data.

The histogram of $\bigl|N_\mathrm{loops}(S_{21})-N_\mathrm{loops}(S_{12})\bigr|$ (shown in Fig.~\ref{GUEGOEGSE-1} (B)) provides a direct topological measure of reciprocity in the scattering matrix. In the GOE class, ordinary TRS forces $S_{21}=S_{12}$, so the loop counts in the forward and reverse channels are virtually identical. As a result, almost all realisations contribute zero difference and the tail of the distribution is very small (any residual nonzero differences can be attributed to finite‐N resolution limits in our loop counting algorithm under computational constraints). Breaking TRS in the GUE ensemble decouples $S_{21}$ from $S_{12}$ so loop counts begin to fluctuate independently, producing a broader distribution with fewer counts at zero and a heavier tail. In GSE simulations, the quaternionic structure and randomised port couplings introduce the strongest departure from simple reciprocity, yielding the lowest density at zero difference and the most populated tail of the three. From these results it is clear that the distributions accurately distinguish between each Gaussian ensemble, and both the central peak height and the tail weight mirrors the degree to which an ensemble respects or violates reciprocity.

\section*{Conclusion}

We have demonstrated that Topological Data Analysis and persistent homology provides useful insights into light-matter interaction by investigating scattering data on a complex plane. The analysis demonstrates that this method is immune to Fano resonance backgrounds that are always present in experimental settings for both strong and weak coupling regimes. The result of TDA is the same even when mode contrast is much lower due to the Fano background effects, and we demonstrated that the method is not equivalent to peak counting as it reveals data features purely encoded in scattering response phase. Additionally, Avoided Level Crossings may be probed with TDA to observe transitions between coupling regimes, and these techniques overall prove to be resilient to the introduction of random trace noise.

These methods are shown to scale to any number of interacting modes, with the response tied to apparent number of degrees of freedom for each particular configuration. Using Random Matrix Theory and the Gaussian ensembles, TDA methods can be further extended to any number of ports and can deal with systems with time reversal symmetry breaking. Ultimately, higher order wave-chaotic systems can be effectively analysed using TDA without relying on spectral transmission data processing methods.

The TDA techniques developed here naturally extend to a range of relevant experimental applications. The ability to analyse nonreciprocal systems may enable accurate probing of exceptional mode coalescence points or the evolution of asymmetric modes. This could expand to photonic systems, where band structure features may be quantified by sampling dispersion surfaces. 

These methods have been shown to be further applicable in systems where high-dimensional spectral data is difficult to analyse with conventional amplitude or phase inspection. Cavity quantum electrodynamic and quantum hybrid systems often present dense modal regimes, ideal situations for the application of the established RMT techniques. Generally, this work finds relevance in any system where parameter-dependent spectral data may be recast in a point cloud representation, enabling the robust extraction of topological features using persistent homology.

This work was supported by the Australian Research Council Grant No. CE110001013.

\hspace{10pt}

\bibliography{biblioBAW}

\end{document}